\documentclass[preprint,showpacs,preprintnumbers,amsmath,amssymb]{revtex4}

\usepackage{epsfig,psfrag}
\usepackage{dcolumn}
\usepackage{bm}
\usepackage{graphicx}
\usepackage{color}
\usepackage{epsfig}
\usepackage{amsmath}
\usepackage{graphicx}
\usepackage{subfigure}

\begin{document}


\title{Counting statistics for the Anderson impurity model: Bethe ansatz
and Fermi liquid study}

\author{A.O.~Gogolin$^1$, R. M.~Konik$^{2}$, A. W. W. Ludwig$^{3}$, and H.~Saleur$^{4,5}$}

\affiliation{${}^{1}$ Department of Mathematics, Imperial College London,
180 Queen's Gate, London SW7 2AZ, United Kingdom \\
${}^{2}$CMPMS Department, Brookhaven National Laboratory, Upton, NY 11973-5000 \\
${}^{3}$Department of Physics, University of California, Santa Barbara, CA 93106 \\
${}^{4}$~Service de Physique Th\'eorique, CEA Saclay, F-91191 Gif-sur-Yvette, France\\
${}^{5}$ Department of Physics, University of Southern California, Los Angeles Ca 90089-0484
}

\date{\today}

\begin{abstract}
We study the counting statistics of charge transport in the
Anderson impurity model (AIM) employing both Keldysh perturbation
theory in a Fermi liquid picture and the Bethe ansatz.
In the Fermi liquid approach, the object of our principal interest is
the generating function for the cumulants of the charge current
distribution.  We derive an \emph{exact analytic formula} relating the full
counting statistic (FCS) generating function to the self-energy of the
system in the presence of a measuring field.  
We first check that our approach reproduces correctly known results 
in simple limits, like the FCS of the resonant level system 
(AIM without Coulomb interaction). We then proceed to study 
the FCS for the AIM perturbatively in the Coulomb interaction.
By comparing this perturbative analysis with a strong coupling
expansion, we arrive at a conjecture for an expression for the FCS
generating function at ${\cal O}(V^3)$ (V is the voltage across the
impurity) valid at all orders in the interaction.

In the second part of the article, we examine a Bethe ansatz analysis
of the current noise for the AIM.  Unlike the Fermi liquid approach, here
the goal is to obtain qualitative, not quantitative, results for a wider
range of voltages both in and out of a magnetic field.  Particularly notable
are finite field results showing a double peaked structure in
the current noise for voltages satisfying $eV \sim \mu_B H$.  This double
peaked structure is the ``smoking gun'' of Kondo physics in the current
noise and is directly analogous to the single peak structure predicted for
the differential conductance of the AIM.

\end{abstract}

\maketitle

\section{Introduction}     \label{introduction}

The subject of counting statistics is rooted in the historical 
paper by Schottky \cite{Schottky} where the measurements of 
charge noise have been carried out and interpreted as the basis
for determining the elementary charge $e$ of the current
carriers: electrons.
Contemporary transport experiments are being performed on
nano-structures, usually involving two electron reservoirs (left
and right) and a central constriction \cite{Goldbaher-Gordon}.
The mean electric current, or linear conductance is well
understood in terms of scattering theory \cite{Landauer} and,
for a single conducting channel, is given by the Landauer
formula:
\begin{equation}\label{Landauerformula}
G_0=\frac{2e^2}{h}T_0\;,
\end{equation}
where $T_0$ is the transmission coefficient and factor 2 stems
from the electron spin. 
However, due to the quantum nature of the problem, the current
is bound to fluctuate. 
In particular, this gives rise to interesting noise
(the second moment of the current distribution) properties
extensively discussed in the literature \cite{Blanter}.

With the third moment of the current distribution now
available experimentally \cite{Reulet}, it is natural
to widen the question to the full current
distribution function or the full counting statistics (FCS).
One way to formulate this question is to ask what is
the probability $P(Q)$ that charge $Q$ will be transmitted
through the system during the waiting time ${\cal T}$ 
and for a given bias voltage $V$.    
As electrons are discrete particles, a naive guess at 
$P(Q)$ would be the Poisson's distribution:
\begin{equation}\label{Poisson}
P(Q)=\frac{\langle Q\rangle^Q}{Q!}e^{-\langle Q \rangle}\;,
\end{equation}
where $\langle Q\rangle=G_0 V{\cal T}$. 
For simplicity, we set $e=\hbar=1$ in what follows. 
The electrons, however, are not only discrete particles
but also quantum particles obeying Fermi-Dirac 
statistics. 
Due to the Pauli principle the electrons will
tunnel `one by one'. 
So, given the `number of attempts',
$N=V{\cal T}/\pi$, one would expect the total probability
be proportional to the probability of successes $T_0^Q$ as
well as the probability of failures $(1-T_0)^{N-Q}$.
The resulting probability distribution is binomial:
\begin{equation}\label{binomial}
P(Q)=\left[\begin{array}{c}Q\\N\end{array}\right] 
T_0^Q(1-T_0)^{N-Q}\;,
\end{equation}
where the binomial coefficient in front simply follows
from the normalization: $\sum_Q P(Q)=1$.
Note that the binomial distribution is a clear signature
of Fermi statistics; indeed, the respective probability
distribution for bosons is the inverse binomial \cite{Mandel}.
In practice, it is more convenient to work with the
generating function $\chi(\lambda)=\sum_Q e^{i\lambda Q}P(Q)$,
where the Fourier transform variable $\lambda$ is
called the `counting field' (see below). The irreducible
moments of the charge distribution immediately follow \cite{Papoulis}: 
\begin{equation}\label{Qmoments}
\ln\chi(\lambda)=\sum\limits_{n=1}^\infty\langle\langle Q^n\rangle\rangle
\frac{(i\lambda)^n}{n!}\;.
\end{equation}
The generating function for the binomial distribution is simply:
\begin{equation}\label{chibinomial}
\chi_{{\rm binomial}}=[1+T_0(e^{i\lambda}-1)]^N\;.
\end{equation}
From this equation one easily recovers the Landauer formula, 
$\langle Q\rangle=NT_0$, the well known expression for
the shot noise $\langle\langle Q^2\rangle\rangle=NT_0(1-T_0)$,
and obtains the following expression for the third moment:
$\langle\langle Q^3\rangle\rangle=NT_0(1-T_0)(1-2T_0)$.
Note that for low transmission ($T_0\to 0$), the statistics reverts
to Poissonian, while for perfect transmission ($T_0\to 1$),
there are no current fluctuations and $\chi(\lambda)=
i\lambda N$. The physics described so far has been understood
in the seminal paper by Levitov and Lesovik \cite{Levitov1} 
(see also \cite{Levitov2}), where they derive a more general
formula for the generating function
\begin{eqnarray}\label{LLformula}
&~&\ln \chi_{0}(\lambda;V;\{T(\omega)\}) = 2 {\cal T} \,
\int  \frac{d \omega}{2 \pi}
\ln \Big\{
1 + T(\omega)\\ &\times&\big[ n_L(1-n_R)
(e^{i \lambda}-1) +
 n_R(1-n_L)(e^{-i \lambda}-1) \big] \Big\} \,,\nonumber
\end{eqnarray}
which is valid for finite voltage, temperatures, and allows
for the energy dependent transmission coefficient. Here
$n_{L/R}(\omega)=n_F(\omega\mp V/2)$ are the thermal electron
distributions in the left and right leads and $n_F(\omega)$ is
the Fermi function.
Clearly $T_0$ is $T(\omega)$ at the Fermi energy, set at $\omega=0$.
Sch\"{o}nhammer has recently re-examined this formula, Eq.(\ref{LLformula}),
by an alternative method, and found that it is correct \cite{Kurt}.

The discussion so far has focused upon {\it non-interacting} electrons.
But while counting statistics for non-interacting electrons
is by now comprehensively understood, the same cannot be said
when the electrons
interact with each other and with the substrate.
Consequently the understanding of the interaction effects on the FCS 
has become an important issue. 
There has been many papers on the
subject in recent years with many interesting yet miscellaneous
results. 
Certainly no general paradigm as to how interactions should
affect the statistics has as yet emerged. 
Misunderstandings dominating the subject only a short time
ago are well illustrated by the following example.
The generating function $\chi(\lambda)$ for Matveev's
Coulomb blockade setup \cite{Matveev} (equivalent to
the $g=1/2$ Kane and Fisher problem \cite{KF}, which is in turn
equivalent to the $\alpha=1/2$ dissipation problem
first solved by Guinea \cite{Guinea})
has been calculated independently 
by three different methods in \cite{AM}, \cite{KT}, and \cite{Komnik1}
with seemingly very different results.
It was only understood later that all three results are 
indeed correct and represent one and the same function
(see Appendix C to Ref. \cite{Gogolin1}).
Moreover the distribution in question turned
out both to be simple and to represent a particular
case of Eqn. (\ref{LLformula}) with a specific choice
of the transmission coefficient, $T(\omega)$.
This lack of a coherent picture of the FCS in strongly correlated systems is,
we believe, simply explained.
As is illustrated by the prominence of the Fermi and Luttinger
liquid paradigms, it is accepted that in the condensed matter it is the {\it low-energy}
physics which is universal.
The FCS is no exception. 
On an energy scale set by the bare parameters, it is therefore the low-temperature,
low-voltage expansions of $\chi(\lambda)$ where universal 
results are to be found. 
The high-voltage (temperature) distributions may be enormously fascinating
but are destined to remain model dependent.

In this paper we collect together 
a number of such universal results, presented
in two parts.  In the first part, Sections II and III, we study
the generating function, $\chi$, using Keldysh perturbation theory
in a Fermi liquid approach.  In Section II,
we introduce the Keldysh method for calculating the 
statistics.  In the process
we establish an exact relationship between the generating function 
and the self energy.
In Section III, we study the FCS for the AIM both
in perturbation theory and in the strong coupling limit.    
By comparing the two we propose a conjecture for $\chi$ at low
voltages, i.e. ${\cal O}(V^3)$, but valid at all orders of the interaction.

In the second part of the paper, Sections IV through VI,
we switch tacts and instead employ a Bethe
ansatz analysis of the current moments for the AIM.  We however limit ourselves to exploring the
behavior of the current and the current noise in the AIM's Kondo regime.  
Our results for these quantities
differ from those of the first section of this paper.  Here the focus is on their
qualitative not quantitative features but over a larger range of voltages (though 
still much smaller
than any bare energy scale) and for finite magnetic fields.  In 
Section IV we review the Bethe Ansatz method
for calculating the current and noise. We then present results for zero
magnetic field in Section V where, in addition, a 
comparison is made to Fermi liquid calculations.
In the final section, Section VI, we consider the properties of the noise in finite
magnetic fields.  Here is found the most significant result of the second part
of the paper.  We are argue that in the vicinity of voltages commensurate with the magnetic field,
the current noise should see a double humped enhancement.  This enhancement is
the analog of that seen in the current when $eV \sim \mu_B H$ and so represents
a 'smoking gun' \cite{winmeir} of Kondo physics.

\section{Keldysh method for the calculation of current statistics:
general results}
\label{SectionI}

The calculation of the charge statistics is usually
accomplished by coupling the system to a `measuring device'. 
In the original gedanken experiment by
Levitov and Lesovik it is a fictitious spin-$1/2$ 
galvanometer coupled to the
current \cite{Levitov2}.
The transmitted charge is then proportional to the change of the
spin phase. As has been shown by Nazarov \cite{Nazarov}, 
the counting of charge can in general be done by coupling 
the system to a fictitious field and
calculating the non-linear response, so leading to 
the same results. In fact the standard quantum mechanical
formula, $P(Q)=\langle \delta(\hat{Q}-Q)\rangle$, can also
be used provided that the central region is initially
decoupled from the leads.

According to \cite{Levitov3} the generating function is given by the
following average,
\begin{equation} \label{chi}
\chi(\lambda)= \left\langle T_{{\rm C}} \exp
\left[-i\int\limits_{{\rm C}} T_\lambda(t)dt \right]
\right\rangle\;,
\end{equation}
where ${\rm C}$ is the Keldysh contour
\cite{Lifshits}, $\lambda(t)$ is the
{\it measuring field} which is non-zero only during the measuring time ${\cal
  T}$:
$\lambda(t)=\lambda\theta (t)\theta({\cal T}-t)$ on the forward path
and $\lambda(t)=-\lambda\theta (t)\theta({\cal T}-t)$ on
the backward path.
Introducing the operator $T_R$ transferring electrons through 
the system in the direction of the current, as well as its
counterpart $T_L$, we can write
\begin{eqnarray}  \label{Tlambdaoperator}
T_\lambda= e^{i\lambda(t)/2}T_R+e^{-i\lambda(t)/2}T_L\;.
\end{eqnarray}
Note that $T_R^\dagger=T_L$
in any system. Consequently, writing out (\ref{chi}) explicitly in 
terms of the time--ordered and anti--time--ordered products, one arrives
at the conjugation property,
\begin{equation} \label{conj}
\chi^*(\lambda)=\chi(-\lambda)\;.
\end{equation}
We now allow $\lambda(t)$ to be an arbitrary function on the Keldysh contour,
$\lambda_\pm(t)$ on the forward/backward path. Then a generalised expression
for the generating function is
\begin{equation} \label{chipm}
  \chi[\lambda_-(t),\lambda_+(t)]=\langle T_{\rm C}
  e^{-i\int_{{\rm C}} \, dt \,
  T_\lambda(t)}\rangle \, .
\end{equation}
Assume that the measuring field changes only very slowly in time.
Then one can write
\begin{equation} \label{potential}
\chi[\lambda_-(t),\lambda_+(t)]=\exp\left[-i\int_0^{\cal T} {\cal
U}[\lambda_-(t),\lambda_+(t)]dt \right],
\end{equation}
where ${\cal U}(\lambda_-,\lambda_+)$ is the {\it adiabatic
potential}. 
Note that Eq.(\ref{potential}) captures the 
leading, linear in ${\cal T}$, contribution to the phase.
For a truly slow-varying measuring field (analytic in $t$), the
corrections are exponentially small in ${\cal T}$.
Even if the function $\lambda(t)$ has isolated jumps,
Eq.(\ref{potential}) still holds at large ${\cal T}$, but
the corrections are then of the order of $\ln{\cal T}$.
This is in full analogy to the non-equilibrium
version \cite{Ng} of the old X-ray problem. 
Once the adiabatic potential is known, the
statistics is given by
\begin{equation}
\ln \chi(\lambda)=-i {\cal T}\, {\cal U}(\lambda,-\lambda)\;.
\end{equation}
Alternatively one can level off the $\lambda_\pm$ functions in
Eq.(\ref{chipm}) to different constants as
\begin{equation}
\chi[\lambda_-(t),\lambda_+(t)]\to\chi(\lambda_-,\lambda_+)\;,
\end{equation}
then
$\chi(\lambda)=\chi(\lambda,-\lambda)$ .
Note that the conjugation property (\ref{conj})
now generalises to
\begin{equation}\label{conj2}
\chi^*(\lambda_-,\lambda_+)=\chi(\lambda_+,\lambda_-)\;,
\end{equation}
or
\begin{equation}
{\cal U}^*(\lambda_-,\lambda_+)=-{\cal U}(\lambda_+,\lambda_-) \, .
\end{equation}

To compute the adiabatic potential we observe that according to the
non-equilibrium version
of the the Feynman--Hellmann theorem \cite{Ng},
\begin{equation}\label{FHel}
\frac{\partial}{\partial\lambda_-}{\cal U}(\lambda_-,\lambda_+)=
\left\langle \frac{\partial T_\lambda(t)}{\partial\lambda_-}
\right\rangle_\lambda\;,
\end{equation}
where the averages are defined as
\begin{equation}
\langle A(t) \rangle_\lambda=\frac{1}{\chi(\lambda_-,\lambda_+)}
\left\langle T_{{\rm C}}\left\{A(t)e^{-i\int\limits_{{\rm C}}
T_\lambda(t)dt} \right\} \right\rangle
\end{equation}
(and similarly for multi--point functions) where $\lambda$'s are
understood to be {\it different} on the two time
branches. Note that the above one--point averages depend on the
branch the time $t$ is on (though not on the value of $t$ on that branch):
\begin{equation}
\langle A(t_-) \rangle_\lambda\neq \langle A(t_+) \rangle_\lambda \, .
\end{equation}
Therefore the average in Eq.~(\ref{FHel}) {\it must} be taken on the
forward branch of the Keldysh contour. 
An advantage of this Hamiltonian approach
is that the calculation of the adiabatic potential ${\cal U}$ 
is reduced to a calculation of Green's functions (GF), 
albeit non-equilibrium ones. So we can use the well developed
diagram technique (and relate to many known results within this
method) without being restricted to the scattering problem.

The conventional test model to consider is the 
Anderson impurity model (see \cite{Hewson} for a review). 
The Hamiltonian of the model consists of three
contributions,
\begin{eqnarray}\label{hamand}
 H = H_0 + H_T + H_C \, .
\end{eqnarray}
The kinetic part
\begin{equation}                \label{kinpart}
 H_0 = \sum_\sigma H_0[\psi_{R/L,\sigma}]+\sum_\sigma(\Delta_0+
\sigma h)d^\dagger_\sigma d_\sigma \, ,
\end{equation}
describes a single fermionic level (which we shall also call the `dot')
with electron creation operators,
$d^\dag_\sigma$ ($\sigma$ is the spin index), energy, $\Delta_0$, and subject
to a local magnetic field, $h$. The electrons in the two non-interacting
metallic leads, $i=R,L$, are represented by $\psi_{R/L, \sigma}$. 
The leads and the dot are coupled via the tunnelling operator,
\begin{equation}                 \label{H_T}
 H_T = \sum_\sigma\left[
\gamma_L e^{i\lambda(t)/2}d^\dagger_\sigma \psi_{L\sigma}
+\gamma_R \psi^\dagger_{R\sigma}
d_\sigma\psi^\dagger_{R\sigma}+{\rm H.c.}\right] \, ,
\end{equation}
with, in general, different amplitudes, $\gamma_{R,L}$. We
have included the counting field into the Hamiltonian (left junction).
One can as well incorporate it into the right junction (or both),
the results are the same due to the gauge symmetry of the
Hamiltonian. Finally, we include the Coulomb repulsion on the dot,
\begin{equation}                 \label{H_C}
 H_C = Un_\uparrow n_\downarrow\; ,
\end{equation}
where $n_\sigma=d^\dagger_\sigma d_\sigma$. The bias voltage $V$ is
incorporated into the full Hamiltonian by assuming different chemical
potentials in the leads: $\mu_L-\mu_R=V\geq 0$.

It is useful to define two auxiliary GFs,
\begin{eqnarray}
 F_\lambda(t,t')
&=& -i\langle T_{{\rm C}}\{\psi_L(t)d^\dagger(t')\}\rangle_\lambda\; \nonumber
 \\ \widetilde{F}_\lambda(t,t')
&=& -i\langle T_{{\rm C}}\{d(t)\psi^\dagger_L(t')\}\rangle_\lambda\; \, .
\end{eqnarray}
Hence the derivative of the adiabatic potential is given by
\begin{eqnarray}                      \label{UF}
\frac{\partial}{\partial\lambda_-}{\cal U}(\lambda_-,\lambda_+)=
\frac{\gamma_L}{2}\lim\limits_{\epsilon\to 0+}
\int\frac{d\omega}{2\pi}e^{i\epsilon\omega}
\nonumber \\ \times
\left[e^{i\lambda_-/2}F^{--}_\lambda(\omega)-e^{-i\lambda_-/2}
\widetilde{F}^{--}_\lambda(\omega)\right]\;.
\end{eqnarray}
As is standard, the mixed GF's can be written as combinations of bare lead
GF's and an exact impurity GF, $D(t,t')$,
\begin{eqnarray}
 \widetilde{F}_\lambda(t,t') &=& \int\limits_{{\rm C}}dt''g_L(t-t'')
 e^{-i \lambda (t'')} D(t'',t') \, , \nonumber \\
F_\lambda(t,t') &=& \int\limits_{{\rm C}}dt''D(t,t')
e^{-i \lambda (t'')} g_L(t''-t')\;.
\end{eqnarray}
Plugging this back into Eq.(\ref{UF}) one obtains
\begin{eqnarray}                   \label{ULR}
\frac{\partial}{\partial\lambda_-}{\cal U}(\lambda_-,\lambda_+)=
\frac{\gamma_L^2}{2} \int \frac{d\omega}{2\pi} \Big[e^{-i
\bar\lambda/2}D^{-+} g^{+-}_L
\nonumber \\
-e^{i \bar\lambda/2}g^{-+}_L D^{+-} \Big] \, ,
\end{eqnarray}
where $\bar\lambda = \lambda_- - \lambda_+$. Thus the
problem is now reduced to calculation of the impurity GF.

To illustrate how the method works we start with the
trivial case of the resonant level model: $U=0$.
Using the GFs of the lead electrons (see, for example, Ref. \cite{Komnik1}),
\begin{eqnarray}   \label{bareGFs}
  g_i^{--}(\omega) &=& g_i^{++}(\omega)=  i 2 \pi \rho_0
 [n_i-1/2] \, , \nonumber \\
 g_i^{-+}(\omega) &=& i 2 \pi \rho_0 n_i \; ,
 \nonumber \\
 g_i^{+-}(\omega)
 &=& - i 2 \pi \rho_0 [1 -  n_i] \; ,
\end{eqnarray}
where $\rho_0$ is the density of states in the electrodes in
the vicinity of Fermi level, one easily obtains the 
bare impurity GF 
(we use the original notation of
Keldysh for the GFs because the standard identity  
$g^{--} +g^{++}=g^{-+}+g^{+-}$ is spoiled by the measuring
field and the Keldysh rotation is useless):
\begin{eqnarray}                    
\hat{D}_0^{-1}(\omega)=
\left[\begin{array}{lr}
\omega-\Delta_0 -i\Gamma_L (2n_L-1)-i\Gamma_R (2n_R-1) &
2ie^{i \bar\lambda/2}\Gamma_L n_L+2i\Gamma_R n_R\\
-2ie^{-i \bar\lambda/2}\Gamma_L (1-n_L)-2i\Gamma_R (1-n_R)&
-\omega+\Delta_0-i\Gamma_L (2n_L-1)-i\Gamma_R (2n_R-1)
\end{array}\nonumber
\right],
\end{eqnarray}
where $\Gamma_{R,L}= (\pi \rho_0 \gamma_{R,L})^2$. 
Inverting this matrix results in 
\begin{eqnarray}
\hat{D}_0(\omega)&=&\frac{1}{{\cal D}_0(\omega)}\label{DbareU}
\nonumber\\
&\times&\left[\begin{array}{lr}
\omega-\Delta_0 +i\Gamma_L (2n_L-1)+i\Gamma_R (2n_R-1) &
2ie^{i \bar\lambda/2}\Gamma_L n_L+2i\Gamma_R n_R\\
-2ie^{-i \bar\lambda/2}\Gamma_L (1-n_L)-2i\Gamma_R (1-n_R)&
-\omega+\Delta_0+i\Gamma_L (2n_L-1)+i\Gamma_R (2n_R-1)
\end{array}
\right]\;,\nonumber
\end{eqnarray}
where $\Gamma = \Gamma_R + \Gamma_L$ and we call the object
\begin{equation}\label{denominatorU=0}
{\cal D}_0(\omega)=(\omega-\Delta_0)^2+\Gamma^2+4\Gamma_L\Gamma_R
\left[n_L(1-n_R)(e^{i \bar\lambda/2}-1)+n_R(1-n_L)(e^{-i
\bar\lambda/2}-1)\right],
\end{equation}
the `Keldysh denominator'. On the technical side, while this
object is a smooth function of energy in the standard
technique (expressible via the advanced and retarded GF's),
here it has discontinuities at the Fermi levels. 
Inserting this result into Eq.(\ref{ULR}) and
integrating over $\lambda_-$ (which can be done in a closed form)
leads to the formula (\ref{LLformula}) with the Breit-Wigner
transmission coefficient,
\begin{eqnarray}
 T(\omega)=\frac{4\Gamma_L\Gamma_R}{(\omega-\Delta_0)^2+\Gamma^2} \, ,
\end{eqnarray}
as expected for the resonant level problem.

Turning to the interacting case, we define the impurity
self-energy in the standard fashion \cite{Lifshits}:
\begin{equation}     \label{selfenergyequation}
 \hat{D}(\omega) = \hat{D}_0(\omega) + \hat{D}(\omega) \hat{\Sigma} (\omega)
\hat{D}_0 (\omega) \, .
\end{equation}
Substituting this into (\ref{ULR}) one obtains a {\it general formula}
for the statistics in interacting one-channel systems
\begin{eqnarray}             \label{fi}
\frac{\partial}{\partial\lambda_-}{\cal U}(\lambda_-,\lambda_+)&=&
-\Gamma_L\int\limits_{-\infty}^\infty\frac{d\omega}{2\pi{\cal
D}(\omega)} \left\{2\Gamma_R\left[e^{i \bar\lambda/2}
n_L(1-n_R)-e^{-i \bar\lambda/2} n_R(1-n_L)\right]
 \right.\\
&-& i \left. \left[ e^{i \bar\lambda/2}n_L\Sigma^{+-}+e^{-i
\bar\lambda/2} (1-n_L)\Sigma^{-+}\right]\right\}\;,\nonumber
\end{eqnarray}
that expresses the generating function in terms of the 
($\lambda$-dependent) impurity self-energy.
Here ${\cal D}(\omega)$ is the
determinant of the (inverse) interacting impurity GF.  
For $\bar\lambda=0$ this equation yields the electric current
and can be shown to reproduce the result by
Meir and Wingreen \cite{Meirwingreen}.
Formula (\ref{fi}) is not restricted to the Anderson model
but is applicable for any similar one-channel systems (for example,
one can add on the electron--phonon interaction or consider
a double dot).

The general formula (\ref{fi}) allows us to investigate
the important limit: linear response statistics at zero temperature. 
Indeed let us take a closer look at the general formula. 
An important technical observation is  that
the limits $V \rightarrow 0$ and $\omega \rightarrow 0$ do not
commute in the presence of the counting field. Indeed, calculating the
Keldysh determinant in both limits we see that even for the 
non-interacting case
\begin{equation}\label{noncom1}
\lim\limits_{\omega\to0}\lim\limits_{V\to0} {\cal
D}_0(\omega,V,\lambda)=\Delta_0^2+\Gamma^2\;,
\end{equation}
but
\begin{equation}\label{noncom2}
\lim\limits_{V\to0}\lim\limits_{\omega\to0} {\cal
D}_0(\omega,V,\lambda)=\Delta_0^2+\Gamma^2+4\Gamma_L\Gamma_R
(e^{i\lambda}-1)\;.
\end{equation}
The latter scheme needs to be implemented when analyzing
the first term in Eq.(\ref{fi}) in the linear response
limit, as the energy integration here is
restricted to $[0,V]$. This leads to a transmission coefficient
type contribution to the generating function (see below).
In the second term in Eq.(\ref{fi}), however, the integration 
over $\omega$ is not restricted to $[0,V]$. 
But due to Auger type effects
\cite{Komnik2} one expects
that there are contributions to the current (and FCS) at all energies.
This effect is itself proportional to the applied voltage 
and only leads to non-linear corrections to the FCS. Hence
the energy integration can in fact be regarded as restricted to $[0,V]$
even in the second term in Eq.(\ref{fi}). Since the
self-energy does not have external lines and all the internal
frequencies have to be integrated over, 
the limits $V \rightarrow 0$ and $\omega
\rightarrow 0$ in this case commute. 
That means that for the evaluation of the
self-energy to the lowest order in $V$ one is allowed to use the
equilibrium GFs, calculated in presence of the counting
field $\lambda$, i.~e. (\ref{DbareU}) with $n_R = n_L = n_F$ and
with the corresponding Keldysh denominator. Therefore
all diagonal Keldysh GFs are equal to those in the equilibrium and
all off-diagonal ones are simply proportional to the same diagrams as
in equilibrium. Since any given off-diagonal self-energy diagram describes
an inelastic process, it should vanish for $\omega \rightarrow 0$ and
we arrive at a conclusion that
\begin{eqnarray}
\lim\limits_{\omega\to 0} \hat{\Sigma}(\omega) = 
\mbox{Re} \, \Sigma^R(0) \left[
 \begin{array}{cc}
  1 & 0 \\
  0 & -1
 \end{array} \right] \, 
\end{eqnarray}
even at finite $\lambda$.
Eq.(\ref{fi}) thus leads to the important result
\begin{eqnarray}          \label{fundformula}
\ln\chi(\lambda)=N\ln\left\{1+\frac{\Gamma^2}{[{\rm
Re}\Sigma^{(R)}(0)]^2 +\Gamma^2}(e^{i\lambda}-1)\right\} \, ,
\end{eqnarray}
or to $\ln\chi(\lambda) = i \lambda N$ for the symmetric Anderson
impurity model. In the case of an \emph{asymmetrically coupled}
impurity, $\Gamma_R \neq \Gamma_L$, the numerator of
(\ref{fundformula}) changes to $\Gamma_R \Gamma_L$ while the
denominator contains $(\Gamma_R + \Gamma_L)/2$ instead of
$\Gamma$.

The result (\ref{fundformula}) allows simple generalisations to
asymmetric systems in a magnetic field $h$. According to
\cite{Yamada} the real part of the self-energy is
given by
\[
{\rm Re}\Sigma^{(R)}_\sigma(0)=\chi_{{\rm
c}}\kappa+\sigma\chi_{{\rm s}}h\;,
\]
where $\chi_{{\rm c/s}}$ are exact charge/spin susceptibilities
and $\kappa\sim\Delta_0+U/2$ is a
particle--hole symmetry breaking field. Consequently
\begin{eqnarray}\label{kondobin}
\ln\chi(\lambda)=\frac{N}{2}\ln\left\{
\left[1+\frac{\Gamma^2}{[\chi_{{\rm c}}\kappa+\chi_{{\rm s}}h]^2
+\Gamma^2}(e^{i\lambda}-1)\right] \nonumber \right. \\
\left. \times \left[1+\frac{\Gamma^2}{[\chi_{{\rm
c}}\kappa-\chi_{{\rm s}}h]^2
+\Gamma^2}(e^{i\lambda}-1)\right]\right\}\;.
\end{eqnarray}
An advantage of this formula is that the
susceptibilities are known \emph{exactly} 
from the Bethe-Ansatz results
\cite{OK,TW}.
We  stress that the result (\ref{fundformula}) is not limited
to the Anderson model but will hold for any similar model, hence the 
{\it binomial theorem}. It is clear in hindsight that all
the non-elastic processes fall out in the $T=0$ linear response
limit. Still it is a remarkable result that {\it all} the 
moments have a simple expression in terms a single number:
the effective transmission coefficient. The binomial distribution
is universal as long as systems with a single conducting channel 
are concerned.

\section{FCS for the Anderson model}
\label{FCSAnderson}

\subsection{Perturbative expansion in the Coulomb interaction}
\label{perttheory}

We now proceed with the  perturbative expansion in the
Coulomb interaction $U$.
The self-energy, up to $U^2$-order, is given, in the time domain, by
\begin{equation}\label{sigma12}
\hat{\Sigma}(t)=\left[
\begin{array}{lr}
-iUD_0^{--}(0) + U^2 [D_0^{--}(t)]^2D_0^{--}(-t) & -U^2
[D_0^{-+}(t)]^2D_0^{+-}(-t) \\
-U^2 [D_0^{+-}(t)]^2D_0^{-+}(-t) & iUD_0^{++}(0)+U^2
[D_0^{++}(t)]^2D_0^{++}(-t)
\end{array}
\right] \, .
\end{equation}
We restrict the calculation to the case of  
the particle-hole symmetric Anderson model $\Delta_0 = -U/2$,.
It can be shown that the contribution to the statistics
linear in $U$ vanishes in the symmetric case. 
We therefore concentrate now on the second-order correction.

The equilibrium self-energy is, in fact, known to all orders 
in $U$ 
\cite{Yamada}:
\begin{equation}\label{sigmaeq}
\hat{\Sigma}_{{\rm eq}}(\omega)=(1-\chi_{{\rm e}})\omega
\left[\begin{array}{lr}1&0\\0&-1\end{array}\right]-
\frac{i\chi_{{\rm o}}^2}{2\Gamma}\omega^2
\left[\begin{array}{ll}{\rm sign}(\omega)&
2\theta(-\omega)\\-2\theta(\omega)&
{\rm sign}(\omega)\end{array}\right]\;,
\end{equation}
where $\chi_{{\rm e/o}}$ are the {\it exact} 
even--odd susceptibilities (i.e.
correlations of $n_\uparrow$ with $n_\uparrow$ and $n_\downarrow$
respectively), which in weak coupling expand as:
\begin{eqnarray}
\chi_{{\rm e}}=1+\left(3-\frac{\pi^2}{4}\right)
\frac{U^2}{\pi^2\Gamma^2}+...\;,\;\;\;
\chi_{{\rm o}}=-\frac{U}{\pi\Gamma}\;.
\end{eqnarray}
For finite $V$ and $\lambda$ in the region $-V/2<\omega<V/2$
we find:
\begin{eqnarray}\label{sigma2}
\hat{\Sigma}(\omega)&=&(1-\chi_{{\rm e}})\omega
\left[\begin{array}{lr}1&0\\0&-1\end{array}\right]\\&-&
\frac{iU^2}{8\pi^2\Gamma^3}
\left[\begin{array}{cc}6\omega V&
e^{-i\lambda}\left(\frac{3V}{2}-\omega\right)^2+
3\left(\frac{V}{2}-\omega\right)^2
\\-e^{-2i\lambda}\left(\frac{3V}{2}+\omega\right)^2-
3e^{-i\lambda}\left(\frac{V}{2}+\omega\right)^2&
6\omega V\end{array}\right]\;.\nonumber
\end{eqnarray}
Further, for $\omega > V/2$ one obtains
\begin{eqnarray}
 \Sigma^{-+}(\omega)=&-& \frac{ie^{-i\lambda}U^2}{8\pi^2\Gamma^3}
\left(\frac{3V}{2}-\omega\right)^2\theta\left(\frac{3V}{2}-\omega\right)
\, ,\label{sigma2b}\\
\end{eqnarray}
while for $\omega < V/2$ the following holds:
\begin{eqnarray}
 \Sigma^{+-}(\omega)=&~&\frac{ie^{-2i\lambda}U^2}{8\pi^2\Gamma^3}
\left(\frac{3V}{2}+\omega\right)^2\label{sigma2bb}
\theta\left(\frac{3V}{2}+\omega\right) \, .
\end{eqnarray}
Substituting these self-energies into (\ref{fi}), 
expanding around the perfect transmission
(hence the sign change of $\lambda$), and formally expressing
the result in terms of the susceptibilities, one finds
the following formula:
\begin{equation}\label{corchi}
\ln \chi(\lambda)=N\left\{i\lambda+\frac{V^2}{3\Gamma^2}
\left[\frac{\chi_{{\rm e}}^2+\chi_{{\rm o}}^2}{4}(e^{-i\lambda}-1)
+\frac{\chi_{{\rm o}}^2}{2}(e^{-2i\lambda}-1)\right]
\right\}+O(V^4)\;.
\end{equation}
This formula is {\it only valid at
the order} $U^2$. There are, however, reasons
to think that it might be exact (see below).
It is tempting to interpret the appearance of the double exponential
terms as an indication of a coherent tunnelling of electron pairs
(caution: similar terms would also appear for the non-interacting 
resonant-level model
due to the energy dependence of the transmission coefficient). 

\subsection{Strong coupling expansion}
\label{strongcoupling}

In the opposite limit of large $U$, the Schrieffer--Wolf type transformation
\cite{Schrieffer}, tailored to the lead geometry \cite{Kaminski},
can be applied and results in a Kondo type model.
For the latter model in the strong--coupling limit, when the
spin on the dot is absorbed into the Fermi sea forming a singlet,
Nozi\`{e}res \cite{Nozieres} devised a Landau--Fermi--liquid description
based on a `molecular field' expansion of the phase shift of the
s--wave electrons:
\begin{equation}\label{phase}
\delta_\sigma(\varepsilon)=\delta_0+\alpha\varepsilon+
\phi^{a}(n_\sigma-n_{\bar{\sigma}}) \;,
\end{equation}
where $\delta_0=\pi/2$, $\alpha$, and $\phi^a$ are phenomenological
parameters corresponding to the residual potential scattering and
the residual interactions, respectively. These processes are
generated by polarizing the Kondo singlet and so are of the order
$\sim 1/T_K$, where $T_K$ is the Kondo temperature. The specific
heat coefficient is proportional to  $\alpha/(\pi\nu)$ while the
magnetic susceptibility is proportional to the sum
$\alpha/(\pi\nu)+2\phi^a/\pi$. Simple arguments were advanced in
Ref.~\cite{Nozieres} to the effect that, because the Kondo
singularity is tied up to the Fermi level, there exists a relation
$\alpha=2\nu\phi^a$ between the two processes in
Eq.~(\ref{phase}). In particular, this explains why the Wilson
ratio is equal to $2$. 
The strong--coupling Hamiltonian that describes the scattering and
interaction processes encoded in Nozi\`{e}res Eq.~(\ref{phase}) is
of the form $H=H_0+H_{{\rm sc}}+ H_{{\rm int}}$. The free
Hamiltonian here is
\begin{equation}\label{Hfree}
H_0=\sum\limits_{p,\sigma}\varepsilon_p(c^\dagger_{p,\sigma}c^{\phantom{\dagger}}_{p,\sigma}
+a^\dagger_{p,\sigma}a^{\phantom{\dagger}}_{p,\sigma})+VQ\;,
\end{equation}
where $c^{\dagger}$ is the creation operator for the s--wave
electrons, $a^{\dagger}$ is the creation operator of the p--wave
electrons, included in order to account for the transport
\cite{Pustilnik}, and the operator
\[
Q=\frac{1}{2}\sum\limits_{p,\sigma}(c^\dagger_{p,\sigma}
a^{\phantom{\dagger}}_{p,\sigma}
+a^\dagger_{p,\sigma}c^{\phantom{\dagger}}_{p,\sigma})\;
\]
stands for the (minus) charge transferred across the junction.
The scattering term is
\begin{equation}\label{Hsc}
H_{{\rm sc}}=\frac{\alpha}{2\pi\nu T_K}\sum\limits_{p,p',\sigma}
(\varepsilon_p+\varepsilon_{p'})c^\dagger_{p,\sigma}c^{\phantom{\dagger}}_{p',\sigma}\;,
\end{equation}
while the interaction term reads
\begin{equation}\label{Hint}
H_{{\rm int}}=\frac{\phi}{\pi\nu^2 T_K}
c^\dagger_{\uparrow}c^{\phantom{\dagger}}_{\uparrow}
c^\dagger_{\downarrow}c^{\phantom{\dagger}}_{\downarrow}\;,
\end{equation}
where $c^{\phantom{\dagger}}_{\sigma}=\sum_pc^{\phantom{\dagger}}_{p,\sigma}$
and we have changed to the dimensionless amplitudes $\alpha$ and $\phi$,
so that in the actual Kondo model $\alpha=\phi=1$ (in the intermediate
calculations it is convenient to treat $\alpha$ and $\phi$ as free parameters
though). By the nature of the strong--coupling fixed point,
the operators $\alpha$ and $\phi$ are irrelevant in the renormalization
group sense and therefore the perturbative expansion in $\alpha$ and $\phi$
is expected to converge.

The shot noise in this model was recently discussed in Refs. \cite{Sela,Golub}.
We now turn to the FCS. To this end we introduce the
the measuring field, which couples, in the Lagrangian
formulation, to the current via a term in the
action $\int dt \lambda(t) \dot{Q}(t)=-\int dt
\dot{\lambda}(t)Q(t)$ that can be gauged away by the canonical
transformation
\begin{equation}\label{rotation}
\begin{array}{llll}
c\to c_\lambda
& = & \cos(\lambda/4) c- i \sin(\lambda/4) a\;,\\
a\to a_\lambda
& = & -i\sin(\lambda/4) c+ \cos(\lambda/4) a\;.
\end{array}
\end{equation}
We therefore reach the conclusion that the charge
measuring field enters this problem as a rotation of the
strong--coupling basis of the s-- and the p--states. While $H_0$
is invariant under this substitution, it should be performed in
both the scattering and the interaction Hamiltonians, $H_{{\rm
sc}}[c]+H_{{\rm int}}[c] \to H_\lambda=H_{{\rm
sc}}[c_\lambda]+H_{{\rm int}}[c_\lambda]$, when calculating the
statistics. It is easily checked that at the first order in
$\lambda$: $H_\lambda=H_{{\rm sc}}+ H_{{\rm int}}
+(\lambda/4)\hat{I}_{{\rm bs}}+O(\lambda^2)$, where $\hat{I}_{{\rm
bs}}$ is the backscattering current operator
\begin{eqnarray}
\hat{I}_{{\rm bs}}=& -&i\frac{\alpha}{4\pi \nu T_K}\sum\limits_{p,p',\sigma}
(\varepsilon_p+\varepsilon_{p'})(c^\dagger_{p,\sigma}a^{\phantom{\dagger}}_{p',\sigma}
-a^\dagger_{p,\sigma}c^{\phantom{\dagger}}_{p',\sigma})\nonumber\\
&-&i\frac{\phi}{2\pi\nu^2 T_K}\sum\limits_{\sigma}
(c^\dagger_{\sigma}a^{\phantom{\dagger}}_{\sigma}-
a^\dagger_{\sigma}c^{\phantom{\dagger}}_{\sigma})
c^\dagger_{\bar{\sigma}}c^{\phantom{\dagger}}_{\bar{\sigma}}\;,
\label{current}
\end{eqnarray}
alternatively available from the commutator
$\hat{I}_{{\rm bs}}=-\dot{Q}=i[Q,H]$.

Applying the standard linked cluster expansion (still valid
on the Keldysh contour, of course) \cite{AGD}, we see that the
leading correction to the distribution function is given by a
connected average
\begin{equation}\label{formalcorr}
\ln\chi(\lambda)=i N \lambda-\frac{1}{2}\int_C dt_1dt_2\langle
T_C\{H_\lambda(t_1)H_\lambda(t_2)\}\rangle_{{\rm c}}+...
\end{equation}
The neglected terms $\alpha^4$, $\alpha^2\phi^2$, $\phi^4$, etc.,
are of the higher order in voltage (temperature) than the main
correction because of the irrelevant nature of the perturbation.
In order to make progress with  Eq.~(\ref{formalcorr}), one only
needs the Green's function of the $\lambda$--rotated
$c$--operator, which is easily seen to be the following matrix in
Keldysh space:
\begin{eqnarray}\label{gfbare}
\hat{g}_\lambda(p,\omega) &=& i\pi\delta(\varepsilon_p-\omega)
\left\{\right.[f(\omega-V/2)+f(\omega-V/2)-1]\hat{\tau}_0\cr\cr
&+& [e^{-i\lambda/2}f(\omega-V/2)+e^{i\lambda/2}f(\omega+V/2)]
\hat{\tau}_+ \cr\cr
&-& [(1-f(\omega-V/2))e^{i\lambda/2}+(1-f(\omega+V/2))e^{-i\lambda/2}]
\hat{\tau}_- \left. \right\}\;,
\end{eqnarray}
where $\hat{\tau}_i$ is the standard choice of Pauli matrices and
$f(\omega)$ is the Fermi distribution function. 

The correction to the distribution function due to the
scattering term (\ref{Hsc}) is: 
\begin{eqnarray}
\delta_\alpha \ln \chi (\lambda) &=&
 -\frac{\alpha^2}{4\pi^2\nu^2
T_K^2}\sum\limits_{p_1,p_2}(\varepsilon_{p_1}+
\varepsilon_{p_2})^2 \int_C dt_1
dt_2 g_{p_2}(t_2,t_1)
g_{p_1}(t_1,t_2)
\nonumber\\
&=& \frac{  \alpha^2{\cal T}}{\pi T_K^2}\int
d\omega\omega^2[(e^{-i\lambda}-1)n_L(1-n_R)+
(e^{i\lambda}-1)n_R(1-n_L)]\;,\label{alphaT}
\end{eqnarray}
which, at zero temperature, contributes to Eq.~(\ref{formalcorr})
a term,
\begin{equation}\label{alphacorr}
\delta_\alpha \ln \chi (\lambda) = \frac{ \alpha^2V^3{\cal
T}}{12\pi T_K^2}(e^{-i\lambda}-1)\;.
\end{equation}

Regarding the correction to the charge distribution coming from
the interaction term (\ref{Hint}), any diagrams with a single
insertion of the Green's function vanish (therefore there is also
no $\alpha\phi$ cross term) and the only remaining connected graph
yields:
\begin{eqnarray}
&& \delta_\phi \ln \chi (\lambda) = -\frac{\phi^2}{2\pi^2\nu^4
T_K^2}\int_C dt_1 dt_2 g(t_1,t_2)^2 g(t_2,t_1)^2
\label{phiT}\\
&=&~\frac{\phi^2}{\pi^2 T_K^2} \int\limits_{-\infty}^{\infty}dt
\frac{\cos^4[\lambda/2+(Vt)/2)]}{(t+i\alpha)^4}
\nonumber\\
&=&~\frac{\phi^2 V^3{\cal T}}{12\pi T_K^2}(e^{-i\lambda}-1)+
\frac{\phi^2 V^3{\cal T}}{6\pi
T_K^2}(e^{-2i\lambda}-1)\;.\nonumber
\end{eqnarray}

Combining the results we find that the zero--temperature
charge distribution function is of the form:
\begin{eqnarray}\label{fullcorr}
\ln\chi(\lambda)&=& i N \lambda+\frac{V^3{\cal T}}{12\pi T_K^2}
\left[(\alpha^2+\phi^2)(e^{-i\lambda}-1)
  \nonumber  \right. \\  &+& \left. 2\phi^2(e^{-2i\lambda}-1)\right]+O(V^5)\;.
\end{eqnarray}

Let us now try to connect these results to the previous weak coupling
calculation. The weak coupling expansion of the susceptibilities 
is given in Section \ref{perttheory}. In the strong coupling limit we have:
\begin{equation}\label{identification}
\chi_e= (\Gamma \alpha)/(\pi T_K)\;\;\;\chi_o=(\Gamma \phi)/(\pi
T_K)\;,
\end{equation}
where in fact $\alpha=\phi=1$ and $T_K$ is the Kondo temperature up to a
pre-factor \cite{TW,OK,Hewson}. The programme of extending a Fermi
liquid approach to non-equilibrium properties of the Anderson
model has not been comprehensively carried out yet.
There is a Fermi-liquid proof, due to Oguri \cite{Oguri}, that the leading
non-equilibrium correction to the zero--temperature current is of the form
\begin{equation}\label{current-Oguri}
I_{{\rm bs}}=\frac{V^3}{12\pi^2\Gamma^2}(\chi_e^2+5\chi_o^2)\;,
\end{equation}
which is valid for all $U$ and interpolates between the
weak--coupling and the strong--coupling regimes of the Anderson
model. We see that the above result for $I_{{\rm bs}}$ is simply the
strong--coupling limit of Oguri's formula.
As to the noise and higher moments
no analogous Fermi--liquid results exist, to the best of our
knowledge. However, we would like put forward a {\it hypothesis}
that Eq.(\ref{corchi}) does represent such a generalisation.
Indeed, we see that this formula is correct up to the $U^2$ order
in weak coupling, it holds in the strong coupling limit,
and it reproduces correctly the mean current at all orders in $U$.

\newcommand{\del}{\partial}
\newcommand{\ep}{\epsilon}
\newcommand{\clsd}{c_{l\sig}^\dagger}
\newcommand{\cls}{c_{l\sig}}
\newcommand{\cesd}{c_{e\sig}^\dagger}
\newcommand{\ces}{c_{e\sig}}
\newcommand{\up}{\uparrow}
\newcommand{\down}{\downarrow}
\newcommand{\il}{\int^{\tilde{Q}}_Q d\la~}
\newcommand{\ilp}{\int^{\tilde{Q}}_Q d\la '}
\newcommand{\ik}{\int^{B}_{-D} dk~}
\newcommand{\ila}{\int d\la~}
\newcommand{\ilpa}{\int d\la '}
\newcommand{\ika}{\int dk~}
\newcommand{\tQ}{\tilde{Q}}
\newcommand{\rh}{\rho_{\rm bulk}}
\newcommand{\ri}{\rho_{\rm imp}}
\newcommand{\sh}{\sig_{\rm bulk}}
\newcommand{\si}{\sig_{\rm imp}}
\newcommand{\rph}{\rho_{p/h}}
\newcommand{\sph}{\sig_{p/h}}
\newcommand{\rp}{\rho_{p}}
\newcommand{\sip}{\sig_{p}}
\newcommand{\drph}{\delta\rho_{p/h}}
\newcommand{\dsph}{\delta\sig_{p/h}}
\newcommand{\drp}{\delta\rho_{p}}
\newcommand{\dsp}{\delta\sig_{p}}
\newcommand{\drh}{\delta\rho_{h}}
\newcommand{\dsh}{\delta\sig_{h}}
\newcommand{\enp}{\ep^+}
\newcommand{\enm}{\ep^-}
\newcommand{\enpm}{\ep^\pm}
\newcommand{\enph}{\ep^+_{\rm bulk}}
\newcommand{\enmh}{\ep^-_{\rm bulk}}
\newcommand{\enpi}{\ep^+_{\rm imp}}
\newcommand{\enmi}{\ep^-_{\rm imp}}
\newcommand{\enh}{\ep_{\rm bulk}}
\newcommand{\eni}{\ep_{\rm imp}}
\newcommand{\sig}{\sigma}
\newcommand{\la}{\lambda}
\newcommand{\om}{\omega}
\newcommand{\no}{\langle I^2\rangle_{\om = 0}}
\newcommand{\dvno}{\partial_V\langle I^2\rangle_{\om = 0}}
\newcommand{\ttk}{{\tilde{T}_k}}

\section{Using the Bethe Ansatz to Compute Current Noise}

In the first part of this paper, we have focused on using Fermi
liquid theory to compute the leading non-trivial correction (i.e. ${\cal O} (V^3)$)
of the generating functional, $\chi (\lambda)$, for both strong and weak coupling.  
We have posited that this computation
is exact and have performed a number of checks indicating that this is so.  
In the second part of the paper we adopt a different tact, 
instead focusing upon general features of the moments of the current (in particular
the noise) in the strong coupling regime over a range of voltages 
(as measured in terms of $T_K$, the Kondo
temperature).  The tool we use to compute the noise 
is the Bethe ansatz.  

\begin{figure}
\begin{center}
\epsfxsize=0.45\textwidth
\epsfbox{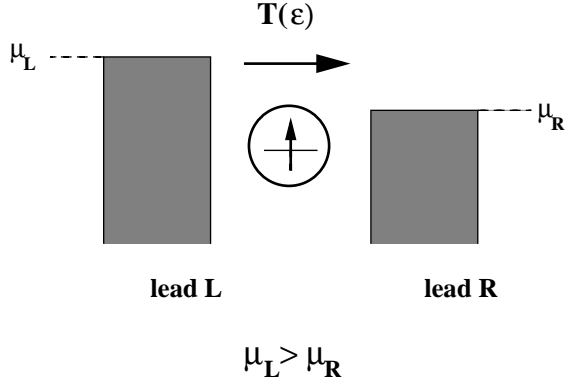}
\end{center}
\caption{A sketch of the distribution of particles in
the leads when $\mu_L > \mu_R$, where $\mu_L$ and $\mu_R$
are the chemical potentials in the two leads.}
\end{figure}
 
Under the Landauer-B\"uttiker formalism \cite{butt}, 
the DC noise, $S$, reduces
to an expression given solely in terms of the transmission 
amplitude, $T_\sigma$, of electronic excitations across the dot:
\begin{equation}\label{eIIIi}
S = \frac{e^2}{h} \sum_\sigma \int^{\mu_L}_{\mu_R}\! d\ep 
~(T_{\sigma} - T_{\sigma}^2) .
\end{equation}
Here we imagine the chemical potential of the left lead is greater than that of the right lead
and so the transmission amplitude, $T_\sigma$, 
governs the flow of electronic scattering states from the left to the right
lead.

The Bethe ansatz gives us the ability to compute $T_\sigma (\epsilon)$ {\it in equilibrium}
up to corrections of ${\cal O}(T_K/\sqrt{U\Gamma})$.  We intend to use these equilibrium
amplitudes to characterize {\it qualitatively} the behaviour of the noise as a function
of $\mu_L-\mu_R$.   
Using equilibrium scattering amplitudes necessarily involves missing some of the
physics present in a non-equilibrium setting.  
Nonetheless we believe that this approach yields important
insights.  In particular, the general behaviour of the noise in a 
magnetic field that arises from this
approach, should be robust enough to survive this particular approximation.  
This is already true
of the current through the dot in a magnetic field where this approach 
\cite{long,moore} yields results consistent with the observed 
enhancement \cite{gold,goldhaber-current}
in the conductance when $\mu_L-\mu_R \sim H$.

\subsection{Computation of the Equilibrium $T_\sigma (\epsilon)$ Using the Bethe Ansatz}

We now turn to a brief description of how to compute $T_\sigma (\epsilon)$ using the Bethe
ansatz solution of the Anderson model.
For additional details see \cite{long}.
The description comes in three parts.
We first describe how to map the {\it equilibrium} (zero voltage) 
problem onto an
one-channel Anderson model,
a model
directly solvable by Bethe ansatz.
Although we work directly in a one lead formulation of the
problem, we are still able to make contact with scattering
in the original two lead picture.
We so identify the relevant elements
of the exact one-channel solution for computing scattering amplitudes.

To reformulate the problem in a way amenable to the Bethe ansatz,
we introduce even/odd electrons
\begin{equation}\label{eIIIii}
\psi_{e/o} = \frac{1}{\sqrt{\gamma_L^2 + \gamma_R^2}} ( \gamma_L \psi_L \pm \gamma_R \psi_R ).
\end{equation}
With this, the odd electron decouples from the dot leaving us with an
interacting theory of even electrons alone:
\begin{eqnarray}\label{eIIIiii}
{\cal H} &=& \sum_{\sigma} \int dx \bigg\{ -i\psi_{e\sigma}^\dagger (x) \del_x \psi_{e\sigma}(x)\cr\cr
&& ~~~~ + (\gamma_L^2+\gamma_R^2)^{1/2} \delta (x) 
( \psi_{e\sigma}^\dagger (x) d_\sigma + d^\dagger_\sigma \psi_{e\sigma} (x) )\bigg\} \cr\cr
&& ~~~~ + \Delta_0 \sum_\sigma n_\sigma + U n_\up n_\down .
\end{eqnarray}
We point out that the Bethe ansatz could be directly applied to the two
channel problem.  However under the diagonalization of $\cal H$ carried out by
the Bethe ansatz, the map (\ref{eIIIii}) is implemented implicitly.
As such we prefer to make the change of basis directly.

Under this one-lead
reformulation, we are still able to make contact with the
scattering amplitudes of electronic excitations off the quantum dot.  Let
$T(\ep)$/$R(\ep)$ be the 
transmission/reflection amplitudes of electronic excitations
of energy, $\ep$, between leads in the original two lead picture.
On the other hand, the even/odd excitations will scatter off the dot with 
some pure phase, $\delta_e(\ep )/\delta_o(\ep)$.  As the scattering in
the odd channel is trivial, $\delta_o(\ep)=0$.  
The two sets
of amplitudes are related according to (\ref{eIIIii}):
\begin{eqnarray}\label{eIIIiv}
e^{i\delta_e(\ep )} &=& R(\ep ) + T(\ep ) ;\cr
e^{i\delta_o(\ep )} &=& 1 =  R(\ep ) - T(\ep ) .
\end{eqnarray}
Our focus will henceforth be on computing $\delta_e$.

To determine $\delta_e (\ep )$, we employ an energetics argument of
the sort used by N. Andrei in the computation of the magnetoresistance
of impurities in a bulk metal \cite{andrei}.
Imagine adding an electron to the system.  
Through periodic boundary conditions,
its momentum is quantized, $p = 2\pi n / L$.  If the dot was absent,
the quantization condition would be determined solely by the conditions
in the bulk of the system and we would write, $p_{\rm bulk} = 2\pi n/L$.
Upon including the dot, this bulk momentum is shifted by a term scaling
as $1/L$.  The quantization condition is then rewritten as
\begin{equation}\label{eIIIv}
p = 2\pi n / L = p_{\rm bulk} + \delta_e(\ep ) / L ,
\end{equation}
where $L$ is the system's length.  
The coefficient of the $1/L$ term is 
identified with the scattering phase of the electron off the dot.
We thus must compute the impurity momenta of excitations in the
problem.

The Bethe ansatz solution of the one channel Anderson model was first
described in \cite{wie} and \cite{kao}.  As with any problem with an $SU(2)$
symmetry, the Bethe ansatz yields a set of quantization conditions describing
two types of excitations, one parameterized by $k$ and associated roughly
with charge excitations, and one parameterized by $\la$ and associated
approximately with spin excitations:
\begin{eqnarray}\label{eIIIvi}
e^{ik_j L + i \delta (k_j)} &=& 
\prod^M_{\alpha = 1} \frac{ g(k_j) - \lambda_\alpha + i/2}
{g(k_j) - \lambda_\alpha - i/2};\cr\cr
\prod^N_{j = 1} \frac{\lambda_\alpha - g(k_j) + i/2}
{\lambda_\alpha - g(k_j) - i/2} &=& - \prod^M_{\beta=1}
\frac{\lambda_\alpha - \lambda_\beta + i}
{\lambda_\alpha - \lambda_\beta - i},
\end{eqnarray}
where 
\begin{eqnarray}\label{eIIIvii}
\delta (k) &=& - 2 \tan^{-1} (\Gamma / (k-\ep_d));\cr\cr
g(k) &=& (k-\ep_d-U/2)^2 / 2 U \Gamma;\cr\cr
\Gamma &=& (V_1^2 + V_2^2).
\end{eqnarray}
Here $N$ is the total number of particles in the system and $M$ marks out
the spin projection of the system, $2S_z = N-2M$ (in zero magnetic field
$M=N/2$).

The possible solutions to the above quantization conditions are manifold.
However most are only significant at finite temperature.  At zero
temperature, the ground state of the system is formed solely from real $k$'s
(and then only if the magnetic field is non-zero) and from bound states
of real $\la$'s together with complex $k$'s.
Specifically, the ground state contains:
\begin{eqnarray}\label{eIIIviii}
&&{\rm i) ~N-2M~real~k_j's} ;\cr\cr
&& {\rm ii) ~M~ real~\lambda_\alpha's};\cr\cr
&& {\rm iii) ~and~associated~with~each~of~the~M~\lambda_\alpha 's~are~}\cr\cr
&& {\rm two~complex~k's,~k^\alpha_\pm ,~described~by~}\cr\cr
&& \hskip .2in g(k^\alpha_\pm) = g(x(\lambda_\alpha) \mp iy(\lambda_\alpha))
= \lambda_\alpha \pm i/2;\cr\cr
&& \hskip .2in x(\lambda ) = U/2 + \ep_d - 
\sqrt{U\Gamma}(\lambda + (\lambda^2+1/4)^{1/2})^{1/2};\cr\cr
&& \hskip .2in y(\lambda ) 
= \sqrt{U\Gamma}(-\lambda + (\lambda^2+1/4)^{1/2})^{1/2}.
\end{eqnarray}
These elementary excitations implement a 
spin-charge separation in the model with the $k$'s representing the charge
sector while the $\lambda$'s represent the spin sector.

In the continuum limit, these excitations are described by
smooth densities, $\rho (k)$ for the real $k$'s and $\sigma (\la )$
for the $\la$'s.  Equations valid at the symmetric
point of the Anderson model describing these
densities can be derived in the standard fashion \cite{wie,kao}:
\begin{eqnarray}\label{eIIIix}
\rho (k) 
&=& \frac{1}{2\pi}+\frac{\Delta (k)}{L}\cr
&& \hskip -.1in 
+ g'(k)\int^\infty_{-\infty} dk'
R(g(k)-g(k'))(\frac{1}{2\pi}+\frac{\Delta(k')}{L})\cr\cr
&& \hskip -.1in 
- g'(k) \int^B_{-D} dk' \rho (k') R(g(k)-g(k'));\cr\cr
\sigma (\la )
&=& \int^\infty_{-\infty} dk
(\frac{\Delta (k)}{L}  + \frac{1}{2\pi})s(\la - g(k)) \cr\cr
&& ~~~~~ - \int^B_{-D} dk \rho (k) s(\la - g(k)).
\end{eqnarray}
where $L$ is the system size and
\begin{eqnarray}\label{eIIIx}
\Delta (k) &=& \del_k \delta (k)/2\pi,\cr\cr
s(x ) &=& \frac{1}{2\cosh (\pi x)}.
\end{eqnarray}
$B$ marks out the `Fermi-surface' of the $k$ 
distribution.  Between $-D$ (the bottom of the band) and $B$ 
there is a sea of $k$ excitations.  The Fermi surface of the $\la$
particles on the other hand is set at $-\infty$: at the symmetric point, the
sea of $\la$ excitations in the ground state
extends from $\la=-\infty$ to $\la=\tilde{Q}$, where $\tilde{Q}$
is the bandwidth of the $\la$ excitations.  
This is a crucial simplification
which makes possible many of our closed-form results.
For most purposes
both bandwidths, $D$ and $\tilde{Q}$, can be taken to be $\infty$.

The density equations neatly
divide into bulk and impurity pieces via
\begin{eqnarray}\label{eIIIxi}
\rho (k) &\rightarrow& \rho_{\rm bulk} (k) + \rho_{\rm imp} (k)/L;\cr\cr
\sigma (\la ) &\rightarrow& \sigma_{\rm bulk} (\la ) + 
\sigma_{\rm imp} (\la )/L .
\end{eqnarray}
The impurity densities of
states contain all the information needed about degrees of freedom
living on the quantum dot.  
The equations governing these densities are
\begin{eqnarray}\label{eIIIxii}
\rho_{\rm imp} (k) 
\!&=&\! \Delta (k)
\!+\! g'(k)\int^\infty_{-\infty}\!\!\!\!\!dk'\! R(g(k)-g(k'))\Delta(k')\cr\cr
&& \hskip -.1in 
- g'(k) \int^B_{-D} dk' \rho_{\rm imp} (k') R(g(k)-g(k'));\cr\cr
\sigma_{\rm imp} (\la )
&=& \int^\infty_{-\infty}dk \Delta (k) s(\la - g(k)) \cr\cr
&& - \int^B_{-D} dk \rho_{\rm imp} (k) s(\la - g(k)).
\end{eqnarray}
For example, the total numbers of
spin $\up$ and $\down$ electrons living on the dot 
are 
\begin{eqnarray}\label{eIIIxiii}
n_{d\up} &=& \int^{\infty}_{-\infty} d\la \sigma_{\rm imp}(\la )
+ \int^B_{-\infty} dk \rho_{\rm imp} (k)\cr\cr
&=& \frac{1}{2} + \frac{1}{2}\int^B_{-\infty} dk \rho_{\rm imp} (k);\cr\cr
n_{d\down} &=& \int^{\infty}_{-\infty} d\la \sigma_{\rm imp}(\la )\cr\cr
&=& \frac{1}{2} - \frac{1}{2}\int^B_{-\infty} dk \rho_{\rm imp} (k).
\end{eqnarray}
The latter equations for each of $n_{d\up}$ and $n_{d\down}$ are
a result of simplifications at the symmetric point.

The energies and momenta of these excitations can be derived
through well known techniques \cite{long}.  The energies are given by
\begin{eqnarray}\label{eIIIxiv}
\ep (k) &=& k - \frac{H}{2} - 2\int d\la ~x(\la ) s(\la - g(k))\cr\cr
&& - \int^B_{-D} dk' g'(k')\ep (k') R(g(k)-g(k'));\cr\cr
\ep (\la ) &=& 2 x(\la ) - 2\int d\la' R(\la -\la')x(\la')\cr\cr
&&+ \int^B_{-D} dk ~g'(k) \ep (k) s(g(k)-\la).
\end{eqnarray}
The momenta are akin to the densities in that they divide
into bulk and impurity pieces \cite{long}.  The bulk momenta are related
directly to the energies via
\begin{eqnarray}\label{eIIIxv}
\ep (k) &=& p(k) - \frac{H}{2};\cr\cr
\ep (\la ) &=& p(\la ). 
\end{eqnarray}
The impurity momenta can be expressed in
terms of the impurity density of states
\begin{eqnarray}\label{eIIIxvi}
\del_k p_{\rm imp}(k) &=& 2\pi \rho_{\rm imp} (k);\cr\cr
\del_\la p_{\rm imp}(\la ) &=& 2\pi \sigma_{\rm imp} (\la ).
\end{eqnarray}
As already discussed, the impurity momenta are the quantities crucial to computing
scattering phases.  These relations will thus allow us to express
the scattering phases in terms of integrals over the impurity density
of states.

In order to determine the scattering phase of an electron (as opposed to
a spin or charge excitation), we 
must specify how to glue together a spin and a charge excitation
to form the electron.  The situation is analogous to adding
a single particle excitation in the attractive Hubbard model.
Adding a single spin $\up$ electron to the system demands that we add
a real $k$ (charge) excitation.  But  
at the same time we create a hole at
some $\la$ in the spin distribution.  The number of the
available slots in the spin distribution is determined by
the total number of electrons in the system.  Adding an electron to the
system thus opens up an additional slot in the $\la$-distribution.
The electron scattering
phase off the impurity is then the difference of the 
right-moving k-impurity
momentum, $p_{\rm imp} (k)$, and the left-moving $\la$-hole 
impurity momentum
$-p_{\rm imp} (\la )$:
\begin{equation}\label{eIIIxvii}
\delta^\up_e = p^\up_{\rm imp} = p_{\rm imp} (k) + p_{\rm imp}(\la ).
\end{equation}
We must now consider how to choose $k$ and $\la$.

As we are interested in the DC noise, we must compute scattering away from
the Fermi surface.  Thus if we are to compute the scattering of an excitation
of energy, $\ep_{el}$, we must choose the $k$ and $\la$ such that
\begin{equation}\label{eIIIxviii}
\ep_{el} = \ep (k) - \ep (\la ).
\end{equation}
However this constraint does not uniquely specify a particular choice
of $(k,\la )$.  We, in general, cannot lift this degeneracy.  However
at the symmetric point of the Anderson model, we can make an ansatz
which has already proven to be successful in the computation of the
finite temperature linear response conductance \cite{long}.
The behaviour of the electron scattering phase is determined by
the impurity densities, $\rho_{\rm imp}$ and $\sigma_{\rm imp}$.  At
the symmetric point of the Anderson model, the scattering phase is
expected to vary as $\sim T_k$, the Kondo temperature.  
Of the two impurity densities, only $\rho_{\rm imp}$ varies as $T_k$ while
$\sigma_{\rm imp}$ is controlled by the much larger scale, $\sqrt{U\Gamma}$.
Thus in computing electronic scattering phases away from the Fermi
surface at $T=0$, it is natural to choose $\la$ at its Fermi surface
value,
i.e. $\la=-\infty$ and vary $k$ according to the energy in which
we are interested.  Specifically, we choose $k$ such that
\begin{equation}\label{eIIIxix}
\ep (k) = \ep_{el}.
\end{equation}
We thus have removed the ambiguity in the choice of $(k,\la )$.

In making this ansatz we are effectively doing the following.  Imagine
an electron in the leads with some energy $\ep_{el}$.  We can imagine
expanding this state in terms of the basis of
Bethe ansatz states:
\begin{equation}\label{eIIIxx}
|el> = \sum_s c_{el,s} |s\rangle ,
\end{equation}
where the states $|s\rangle$ are {\it exact} 
eigenfunctions of the Hamiltonian.
Although we only possess incomplete knowledge of this expansion, it
would be a reasonable guess that it contains multiple
terms.  However our ansatz supposes only a single state 
contributes.  But because of the hierarchy of scales, 
$T_k \ll \sqrt{U\Gamma}$,
in the problem, we expect additional terms in the expansion of
Eqn. (\ref{eIIIxx}) to have coefficients 
of ${\cal O}(T_k/\sqrt{U\Gamma})$.

Under this ansatz, the scattering phase of the spin $\up$ electron
at some energy, $\ep_{el}$, above the Fermi surface is then
\begin{eqnarray}\label{eIIIxxi}
\delta_e^{\up}(\ep_{el} ) &=&  p_{\rm imp} (k) + p_{\rm imp}(\la=-\infty );\cr\cr
&=& 2\pi\int^{\tilde{Q}}_Q d\la \sigma_{\rm imp} (\la ) 
+ 2\pi\int^k_{-D} dk' \rho_{\rm imp} (k') ,\cr\cr
&& \hskip 1.5in  ~~~\ep (k) = \ep_{el} .
\end{eqnarray}
When the magnetic field, $H$, is 0, $\ep (k) >0$ and we can only
directly compute the scattering of spin $\up$ electrons.  However
with $H>0$, $\ep (k)$ takes on negative values and so we can also compute
spin $\up$ hole scattering.  To add a spin $\up$ hole with energy,
$\ep_{hole} > 0$, we remove a $k$ and a $\la$-hole in the spirit
of our previous ansatz:
\begin{eqnarray}\label{eIIIxxii}
\ep (k) &=& -\ep_{hole};\cr\cr
\lambda &=& -\infty .
\end{eqnarray}
The scattering phase is then
\begin{eqnarray}\label{eIIIxxiii}
\delta_{ho}^{\up}(\ep_{hole} ) &=&  p_{\rm imp} (k) + p_{\rm imp}(\la )\cr\cr
&=& 2\pi\int^{\tilde{Q}}_Q d\la \sigma_{\rm imp} (\la ) 
+ 2\pi\int^k_{-D} dk' \rho_{\rm imp} (k') ,\cr\cr
&& \hskip 1.35in  ~~~\ep (k) = -\ep_{hole} .
\end{eqnarray}
So far we have computed the scattering of spin $\up$ objects.  To
compute spin $\down$ quantities, we employ the particle-hole transformation
relating spin $\up$ to spin $\down$:
\begin{eqnarray}\label{eIIIxxiv}
c^\dagger_\up (k) &\rightarrow& c_\down (-k) ;\cr\cr
c^\dagger_\down (k) &\rightarrow& c_\up (-k) ;\cr\cr
d^\dagger_\up &\rightarrow& d_\down ;\cr\cr
d^\dagger_\down &\rightarrow& d_\up ;\cr\cr
\ep_d & \rightarrow& \ep_d ,
\end{eqnarray}
where the last line only follows at the symmetric point.
With this transformation, we obtain
\begin{eqnarray}\label{eIIIxxv}
{\delta^\down}_{el}(\ep_{el}) &=& \delta^\up_{ho}(\ep_{ho}=\ep_{el});\cr\cr
{\delta^\down}_{ho}(\ep_{ho}) &=& \delta^\up_{el}(\ep_{el}=\ep_{ho}).
\end{eqnarray}
Our inability to directly compute spin $\down$ scattering is a
technical peculiarity of the Bethe ansatz \cite{long}.

\subsection{Nature of the Approximation}

We have now described how to compute the scattering amplitudes
as a function of energy in equilibrium.  The error in using these amplitudes
in describing out-of-equilibrium quantities such as the current or the noise
has two possible sources.  The first source can be seen from the way the
differing chemical potentials of the right and left reservoirs appear in the Hamiltonian:
\begin{equation}\label{eIIIxxvi}
{\cal H}_\mu = \mu_L \int dx~\psi^\dagger_L (x) \psi_L (x) + 
\mu_R \int dx~\psi^\dagger_R (x) \psi_R (x).
\end{equation}
Under the map to the even and odd basis (\ref{eIIIii}), this term becomes non-diagonal,
coupling the even and odd sectors.  This coupling between sectors, in turn, lifts the model's
integrability, 
the basis on which we compute $T_\sigma(\epsilon)$.  One might have hoped \cite{long} 
that the state of the system in
non-equilibrium could still be characterized by using in-equilibrium data in a fashion 
analogous to the manner
in which out-of-equilibrium quantum Hall edges can be characterized exactly \cite{fendley}.  
While in analyzing the Hall edges, an even-odd transformation of bosonic degrees of freedom is employed, the
Hall case is much simpler as the current as well as the coupling to the voltage bias, can be expressed
directly in terms of the odd degree of freedom.  
But in the case of the Anderson model,
the current and voltage involve both even and odd degrees of freedom.  In mapping back to the
left-right basis, 
some sort of breaking of integrability then occurs.  We will see this explicitly when we compute the
noise at zero magnetic field and compare it to the exact Fermi liquid results.

A second source of error in using the equilibrium scattering amplitudes 
concerns the manner in which we construct
the electronic scattering states.  As we have already indicated, there is a 
multiplicity of choices in how we
construct the scattering states.  The particular choice we employed was the 
simplest that met the requirement that the energy
dependence of the scattering varies on the Kondo scale.  However this choice is 
not unique and while it produced
excellent results for the behaviour of the finite temperature linear response 
conductance, it may not be optimal
for the description of out-of-equilibrium scattering.

\section{Computation of Noise in Zero Field: Comparison between Fermi Liquid Theory and Bethe Ansatz }

In this section we compute the noise in zero magnetic field and compare it
at small voltages to the Fermi liquid results from the first part of this article.
To compute the noise, we imagine biasing the leads as in Figure 1 with $\mu_R < \mu_L$.
For convenience we set $\mu_L$ to zero and $\mu_R \equiv \mu$.
Then as discussed in the previous section, the noise is given by
\begin{equation}\label{eIVi}
S = \frac{e^2}{h} \int^0_{\mu} 
d\ep \big(T_\up (1-T_\up) + T_\down (1-T_\down )\big).
\end{equation}
At $H=0$ spin $\up$ and spin $\down$ scattering are the same,
i.e. $T_\down = T_\up$.  As we are scattering electrons from lead $R$ 
with energy $\ep < 0$ into lead $L$, we are equivalently interested in computing 
the scattering phase of a hole.
In Section IV we demonstrated that at $H=0$
we are able to compute the 
scattering amplitude of a spin $\down$ hole.  Specifically we have
\begin{eqnarray}\label{eIVii}
T_\down (\ep ) &=& T_\up (\ep ) = 
\sin^2 (\frac{\delta^\down_{ho}(\ep )}{2});\cr
\delta^\down_{ho} (\ep ) 
&=& 2\pi\int^{\tilde{Q}}_Q d\la \sigma_{\rm imp} (\la ) 
+ 2\pi \int^k_{-D} dk' \rho_{\rm imp} (k'), \cr\cr
&& \hskip 1.5in ~~~  \ep(k) = \ep.
\end{eqnarray}
In \cite{long} we were able to evaluate these expressions
in closed form:
\begin{eqnarray}\label{eIViii}
\delta^\up_{ho} (\ep ) &=&  \frac{3}{2}\pi - \sin^{-1} 
\big(\frac{1-\ep^2/\tilde{T}_k^2}{1+\ep^2/\tilde{T}_k^2}\big)\cr\cr
&&\hskip -.15in  + 2\sum^\infty_{n=0} \frac{1}{1+2n} 
\big(\frac{\ep \pi}{\sqrt{2U\Gamma}}\big)^{1+2n}\cr\cr
&&\hskip -.05in \times\int dk e^{-\pi g(k)(1+2n)} {\rm Re}[\Delta (ik)];\cr\cr
\tilde{T}_k &\equiv& \frac{2}{\pi}T_k = \frac{2}{\pi}\sqrt{\frac{U\Gamma}{2}}
e^{-\pi (\frac{U}{8\Gamma} - \frac{\Gamma}{2U})}.
\end{eqnarray}
The last equation gives the crossover scale, $T_k$, the Kondo
temperature, in terms of the bare parameters of the model \cite{haldane}.
With this, $T_{\up/\down}$ equals
\begin{equation}\label{eIViv}
T_{\up/\down} = \frac{1}{1+\frac{\ep^2}{\tilde{T}_k^2}} + 
{\cal O}(\frac{T_k}{\sqrt{U\Gamma}}).
\end{equation}
For typical realization of Kondo physics in quantum dots, i.e. \cite{gold,cron}, the error term is insignificant.
And so we compute the noise to be
\begin{equation}\label{eIVv}
S (\mu ) = \frac{e^2}{h}\bigg( \frac{\mu}{1 + \frac{\mu^2}{\ttk^2}}
- \ttk \tan^{-1} (\frac{\mu}{\ttk})\bigg).
\end{equation}
The quantity $S / V$ is solely a function of the ratio $V/T_k$ and is
plotted against $I/|V|$ in Fig. 2.  (Note that $\mu = eV$.)
We see that as $V/T_k$ is varied,
the noise rapidly rises at first, peaks at $eV \approx -1.15 T_k$, and then
begins to gradually decline.

This behaviour is closely related to the Kondo resonance in the spectral
impurity density of states.  As we express the scattering phases in terms
of the impurity density of states, we probe the resonance.  
The width, $w$, of this resonance, $w\sim T_k $, corresponds to the energy scale
at which we expect maximal noise.

The noise in this case can be reexpressed in terms of the current, $I$, and the
differential conductance, $G$,
\begin{eqnarray}\label{eIVvi}
I(\mu ) &=& -2\frac{e}{h} \ttk \tan^{-1} (\frac{\mu}{\ttk});\cr\cr
G(\mu ) &=& 2\frac{e^2}{h} \frac{1}{1+ \frac{\mu^2}{\ttk^2}},
\end{eqnarray}
with the result
\begin{equation}\label{eIVvii}
S(\mu ) = \frac{1}{2} \mu G(\mu ) + \frac{e}{2} I(\mu ).
\end{equation}
At small $\mu$, 
$$
S(\mu < 0) = -\frac{2e^2}{h} \frac{8\pi^2}{96}{\mu^3}{T_K^2}.
$$
We can compare this result with the Fermi liquid result
$$
S^{FL}(\mu) = -\frac{2e^2}{h} \frac{5\pi^2}{96}{\mu^3}{T_K^2}.
$$
In making this comparison, there is a certain arbitrariness in how one defines
$T_K$.  This can be overcome by appealing to the finite temperature 
linear response conductance,
$G(T)$ to fix the manner in which $T_K$ is to be defined.  In our conventions then, 
$G(T) = 2e^2/h(1-\frac{\pi^4}{16}\frac{T^2}{T_K^2}$).
We see then that the Fermi liquid result differs from the result based upon the equilibrium
Bethe ansatz scattering states by a factor of 5/8.

A similar difference can be found between the Fermi liquid and Bethe ansatz computation 
for the current.
At small $\mu$, the leading order correction to the current (of ${\cal O}(\mu^3)$) is given by
\begin{equation}\label{eIVviii}
\delta I(\mu )  = 2\frac{e}{h}\frac{\pi^2}{12}\frac{\mu^3}{T_K^2}.
\end{equation}
This compares to the Fermi liquid result
\begin{equation}\label{eIVix}
\delta I^{FL}(\mu )  = 2\frac{e}{h}\frac{\pi^2}{32}\frac{\mu^3}{T_K^2}.
\end{equation}
We see that the Fermi liquid result is considerably smaller 
than that of the Bethe ansatz.  This might
well reflect the role of incoherent scattering processes that would be unaccounted for properly by
using equilibrium scattering amplitudes.

Finally we consider the value of the effective charge, $e^*$, in the problem.   
This charge is given as a ratio
of the noise to the backscattering current:
\begin{equation}\label{eIVx}
e^* = S/I_{bs}.
\end{equation}
In the case of the Bethe ansatz, we find $e^* = e$, that is, we find the Johnson-Nyquist result for
shot noise in the weak scattering limit.  However in Fermi liquid theory, the effective
charge is found to be $e^{*FL} = 5/3 e$ \cite{Sela}.  That the effective charge goes unchanged 
from its non-interacting
value is again presumably a consequence of the use of equilibrium scattering states.

While the use of the equilibrium Bethe ansatz scattering states gets quantitative details of the noise 
(and the current) incorrect, it gets qualitative features correct.  In particular the conductance
as a function of voltage is Lorenztian-like with a width of order $T_K$.  Similarly the zero field
noise as a function of voltage increases rapidly (on a scale of order $T_K$) and thereafter decreases 
slowly.  Both of these features would be expected to be present based solely upon the presence of the Kondo/Abrikosov-Suhl
resonance whose width is governed by the scale $T_K$.

\begin{figure}
\begin{center}
\epsfxsize=0.45\textwidth
\epsfbox{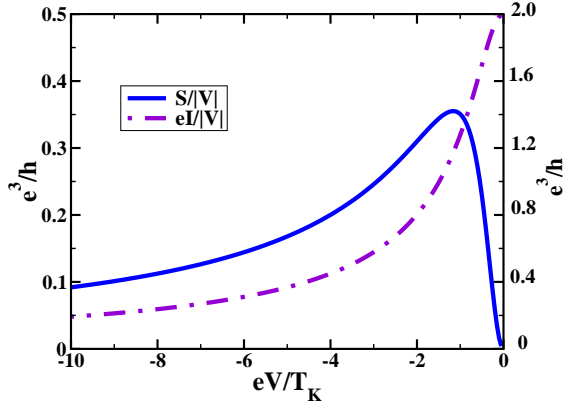}
\end{center}
\caption{Plot describing the evolution
of both the noise, $S/V$, and
the current, $I/V$, as a function of the applied voltage. 
The scale on the l.h.s. governs the noise while the scale on
the r.h.s. governs the current.}
\end{figure}

\section{Computation of Noise in Finite Field} 

In this section we compute the noise in a finite magnetic (Zeeman) field.  
We will here argue that the finite field
noise possesses features that can be considered a
``smoking gun'' \cite{winmeir} of Kondo physics.  We believe that these 
features are robust and so should be captured
by our approach.

To compute the noise,
\begin{equation}\label{eVi}
S = \frac{e^2}{h} \int^0_{\mu} d\ep 
\big(T_\up (1-T_\up) + T_\down (1-T_\down )\big),
\end{equation}
in a finite field, we must consider the contributions of the spin $\up$ and
spin $\down$ currents individually.  
From Section III, $T_\up$ and $T_\down$ are
given by
\begin{eqnarray}\label{eVii}
T_\up &=& \sin^2 (\frac{\delta^\up_{ho} (\ep )}{2});\cr\cr
\delta^\up_{ho} (\ep ) &=& 2\pi\int^{\tilde{Q}}_Q d\la \sigma_{\rm imp} (\la ) 
+ 2\pi \int^k_{-D} dk' \rho_{\rm imp} (k'), \cr\cr
&& \hskip 1.5in \ep(k) = -\ep;\cr\cr
T_\down &=& \sin^2 (\frac{\delta^\down_{ho} (\ep )}{2});\cr\cr
\delta^\down_{ho} (\ep ) &=& 2\pi\int^{\tilde{Q}}_Q d\la \sigma_{\rm imp} (\la ) 
+ 2\pi \int^k_{-D} dk' \rho_{\rm imp} (k'), \cr\cr
&& \hskip 1.5in  \ep(k) = \ep.
\end{eqnarray}
In \cite{long} we evaluated these expressions in two cases, $H \ll T_k$ and
$H > T_k$:

\vskip .3in 

\noindent {\bf case i: $H \ll T_k$}
\vskip .2in
We found for spin $\up$ hole scattering
\begin{eqnarray}\label{eViii}
\delta_{ho}^{\up}(\ep_{ho}>0) &=& \frac{5}{4}\pi -
\sin^{-1}\big(\frac{1-(\ep_{ho} - H)^2/\tilde{T}_k^2}{1+ (\ep_{ho} - H)^2/{\tilde{T}_k}^2}\big)\cr\cr 
&& + \frac{1}{2}\sin^{-1}\big(\frac{1-H^2/{\tilde{T}_k}^2}{1+ H^2/{\tilde{T}_k}^2}\big), 
\end{eqnarray}
while for spin $\down$ hole scattering we arrived at
\begin{eqnarray}\label{eViv}
\delta_{ho}^{\do}(\ep_{ho}>0) &=& \frac{5}{4}\pi - 
\sin^{-1}\bigg(\frac{1-(\ep_{ho} + H/2)^2/{\tilde{T}_k}^2}
{1 + (\ep_{ho} + H/2)^2/{\tilde{T}_k}^2}\bigg)\cr\cr
&& + \frac{1}{2}\sin^{-1}\bigg(\frac{1-H^2/(4{\tilde{T}_k}^2)}
{1+ H^2/(4{\tilde{T}_k}^2)}\big) .
\end{eqnarray}
Consequently the transmission amplitudes in this case equal
\begin{eqnarray}\label{eVv}
T^\up &=&  \frac{1}{2}\bigg(1 + \frac{1+(H^2-\mu_2^2)/{\tilde{T}_k}^2}
{(1+H^2/{\tilde{T}_k}^2 )^{1/2}(1+(\mu_2+H)^2/{\tilde{T}_k}^2)}\bigg);\cr\cr\cr
T^\down &=& \frac{1}{2}\bigg(1+
\frac{1+({H^2/4}-\mu_2^2)/{\tilde{T}_k}^2}
{(1+H^2/4{\tilde{T}_k}^2 )^{1/2}
(1+(\mu_2-H/2)^2/{\tilde{T}_k}^2)}\bigg).
\end{eqnarray}
\vskip .3in

\noindent {\bf case ii: $H>T_k$}
\vskip .2in

Using a Weiner-Hopf analysis, $\delta^\up_{ho}$ and $\delta^\down_{ho}$ 
were determined in Ref. \cite{long} to be 
\begin{eqnarray}\label{eVvi}
\delta_{ho}^\up &=& \pi + 2\tan^{-1}(2(I^{-1}-g(k)));\cr\cr
\delta_{ho}^\down &=& \frac{3\pi}{2} + \tan^{-1}(2(I^{-1}-b)),
\end{eqnarray}
where $I^{-1}$ sets the Kondo scale, $T_k \sim e^{-\pi I^{-1}}$, and
$I^{-1}-b$ is given in terms of the ratio $H/T_k$:
\begin{eqnarray}\label{eVvii}
I^{-1} &=&  \frac{U}{8\Gamma} - \frac{\Gamma}{2 U};\cr\cr
I^{-1}-b &=& \frac{1}{\pi}\log \big(\frac{H}{2T_k}\sqrt{\frac{\pi e}{2}}\big).
\end{eqnarray}
$k$ is parameterized in terms of the energy, $\ep$, by the expression

\begin{eqnarray}\label{eVviii}
\ep(k) &=& -H\bigg(1 - \frac{1}{2\pi} \tan^{-1}\frac{1}{g(k)-b}\cr\cr
&& ~~~~~~~~- \frac{1}{4\pi^2}\frac{1}{1 + (g(k)-b)^2}
\bigg[ \frac{\psi(1/2)}{\Gamma (1/2)} + 1\
- (g(k)-b)\tan^{-1}(\frac{1}{g(k)-b}) \cr\cr
&&\hskip 1.8in + {\bf C} + 
 \frac{1}{2}\log (4\pi^2(1+(g(k)-b)^2))\bigg]\bigg)\cr\cr
&&+ \frac{\sqrt{2\Gamma U}}{\pi^2} \bigg(\frac{1}{\sqrt{2e\pi}}
\frac{e^{-b\pi}}{1 + (g(k)-b)^2} 
+ e^{-\pi g(k)}\tan^{-1}(\frac{1}{g(k)-b})\bigg)\cr\cr
&&+ {\cal O} ((g(k)-b)^{-3}) 
\end{eqnarray}

\noindent where $C=.577216\ldots$ is Euler's constant
and $b$ is given by
\begin{equation}\label{eVix}
b = \frac{1}{\pi} \log (\frac{2}{H}\sqrt{\frac{U\Gamma}{\pi e}}).
\end{equation}
With this 
\begin{eqnarray}\label{eVx}
T^\up &=& \frac{1}{1+(2(I^{-1}-g(k))^2)} + \cdots;\cr\cr
T^\down &=& \frac{1}{2} - \frac{I^{-1}-b}{(1+4(I^{-1}-b)^2)^{1/2}} + \cdots.
\end{eqnarray}
In writing (\ref{eVx}) we have omitted writing terms arising from
the full expression for $\ep (k)$ in (\ref{eVviii}).  But
because of the logarithmic dependence upon $H/T_k$,
such terms are needed if we are to compute the noise with reasonable
accuracy for fields, $H$, not far in excess of $T_k$.
Notice that the spin $\down$ scattering does not vary as a function
of energy, an approximation valid for $H \gg T_k$.

\begin{figure}
\begin{center}
\epsfxsize=0.45\textwidth
\epsfbox{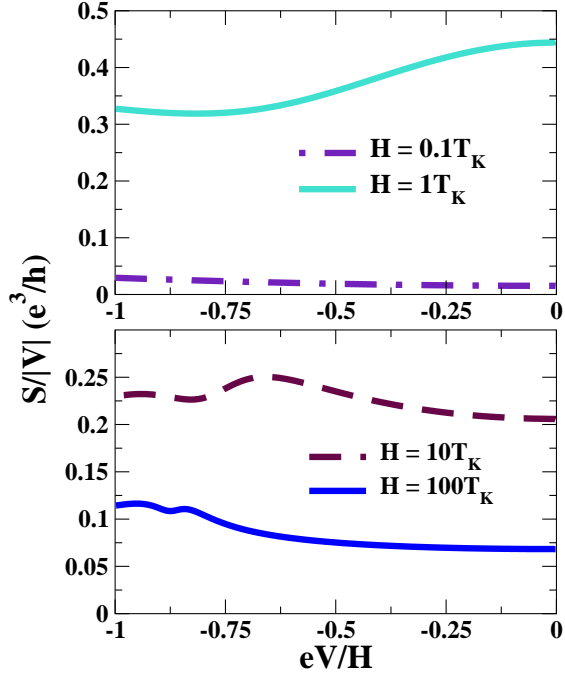}
\end{center}
\caption{Plots describing the behaviour
of the noise, $S/V$,
as a function of the applied voltage, $V$, for a variety of magnetic
field values.}
\end{figure}

We plot the noise, $S/|V|$, in Figure 3 for a variety of values of $H/T_k$.
For small $H/T_k$ the noise is smooth and without structure.
However as we vary $H$ from $H<T_k$ to $H>T_k$, a marked variation occurs
as seen in top panel of Figure 3.  The noise rapidly increases, achieving
its maximal value of roughly $H\sim T_k$, the crossover scale, before
again decreasing.  This behaviour is repeated in the differential noise $-\partial_VS$,
plotted in Figure 4.  The noise's maximal value at $H=T_k$ is a reflection
of the maximum in $T^\up(1-T^\up)$.  
Such a maximum occurs for $T^\up \sim 1/2$.
Thus at $H=T_k$, the average transmission amplitude for spin $\up$
excitations has been reduced roughly
by $1/2$.

At large $H/T_k$, the noise develops a double humped structure near $e|V|\sim H$.
This feature is more apparent when we examine the differential noise in
the lower panel of Figure 4.  In varying $V$ near
the peak, the differential conductance $G\propto T^\up (V)$, passes through
the value $G_{max}/2=e^2/(2h)$ twice.  As such the quantity,
$\partial_V S \sim T^\up (1-T^\up )$, possesses two peaks.

Given the bias at which it occurs,
the doubled peak is intimately related to the
peak in the differential conductance seen near $eV \sim H$.
The peak in the differential conductance owes its origin to a
field induced bifurcation in the Kondo resonance \cite{winmeir}: as the Kondo
resonance shifts so does the peak in the conductance.
In Refs. \cite{long,moore} this bifurcation was studied where it was found that
the peak occurs at a 
value of $eV$ distinctly smaller than $H$ and not $eV=H$.
In the case of the noise, we again find that the peaked structure in
it occurs at values of $eV$ smaller than $H$.

We believe this double peaked structure in the noise, inasmuch as it depends on
the gross dependency of the scattering amplitudes upon H, to be a robust feature.
Less certain are the quantitative predictions that arise from this analysis.
Nonetheless we will proceed to analyze the structure of the differential noise peaks.  
We do point out that we 
have some confidence that this analysis has merit as a subset of 
its corresponding predictions
for the behavior of the {\it current} in a magnetic field have been shown to be 
at least qualitatively correct \cite{goldhaber-current}.  In particular, the
prediction that the peak in the differential conductance 
occurs for values of $eV$ smaller than $H$ has been observed in experiments
on carbon nanotubes \cite{goldhaber-current}.

The differential noise, $-\partial_VS$, is given by
\begin{equation}\label{eVxi}
\partial_V S = \frac{e^3}{h} \big( T^\up (1-T^\up ) + T^\down (1-T^\down )\big) .
\end{equation}
For $V \sim H$ and $H \gg T_k$, $T_\down \ll T_\up$, the locations of
the peaks again occur when
$$
T^\up = 1/2 .
$$
As $T^\up = \sin^2(\delta^\up_{ho}/2)$, the
scattering phases that correspond to this amplitude are, 
$\delta^{peak\up}_{ho} = \frac{\pi}{2} / \frac{3\pi}{2}$.  From (\ref{eVvi}),
this in turn implies 
$g(k) = I^{-1} \pm 1/2$.  Using \ref{eVviii}, the biases, $V_\pm$,
at which the two peaks occur equal
\begin{equation}\label{eVxii}
eV_\pm = -\frac{H}{2\pi} \frac{1}{I^{-1} - b \pm \frac{1}{2}}.
\end{equation}

\begin{figure}
\begin{center}
\epsfxsize=0.45\textwidth
\epsfbox{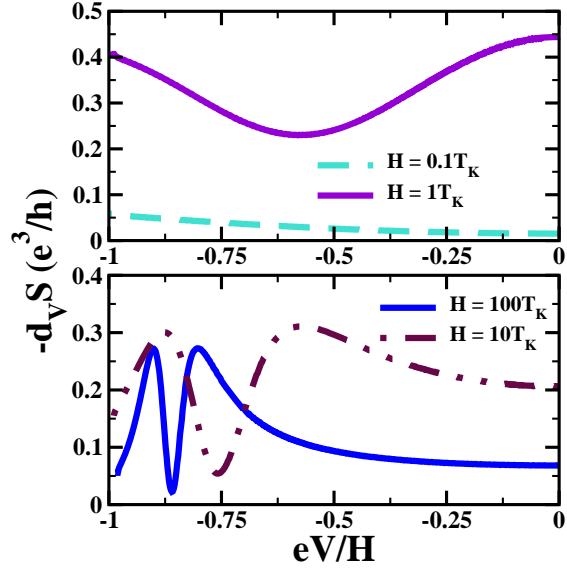}
\end{center}
\caption{Plots describing the behaviour
of the differential noise, -$\partial_V S$,
as a function of the applied voltage, $V$, for a variety of magnetic
field values.}
\end{figure}

The height of the peaks can also be determined.  The maxima of the peaks
occur when $T^\up = 1/2$.  Consequently, the height of the peaks in
$-\partial_V S$ are given by adding $e^3/(4h)$, the contribution due to the
spin $\up$ current, to the contribution from spin $\down$ scattering
with the result,
\begin{equation}\label{eVxiii}
-\partial_V S^{\rm max} = \frac{e^3}{4h} \frac{4(I^{-1}-b)^2 +2}{4(I^{-1}-b)^2 +1}.
\end{equation}
This result holds for either peak, a consequence of the lack of variation
in $T^\down$ for $eV,H \gg T_k$.

We are also able to compute the full width at 
half maximum (FWHM) of the peaks.
As the peak maxima occur for $T^\up(1-T^\up) = 1/4$,
or phases, $\delta^{peak\up}_{ho} = \frac{\pi}{2}/\frac{3\pi}{2}$,
the FWHM occur for $T^\up (1-T^\up) = 1/8$.
For the peak, $\delta^{peak\up}_{ho} = \pi/2$, 
the phases of the FWHM then correspond
to 
$$
\delta^{FWHM\up}_{ho} = \frac{\pi}{4},~\frac{3\pi}{4}.
$$

\begin{figure}
\begin{center}
\epsfxsize=0.45\textwidth
\epsfbox{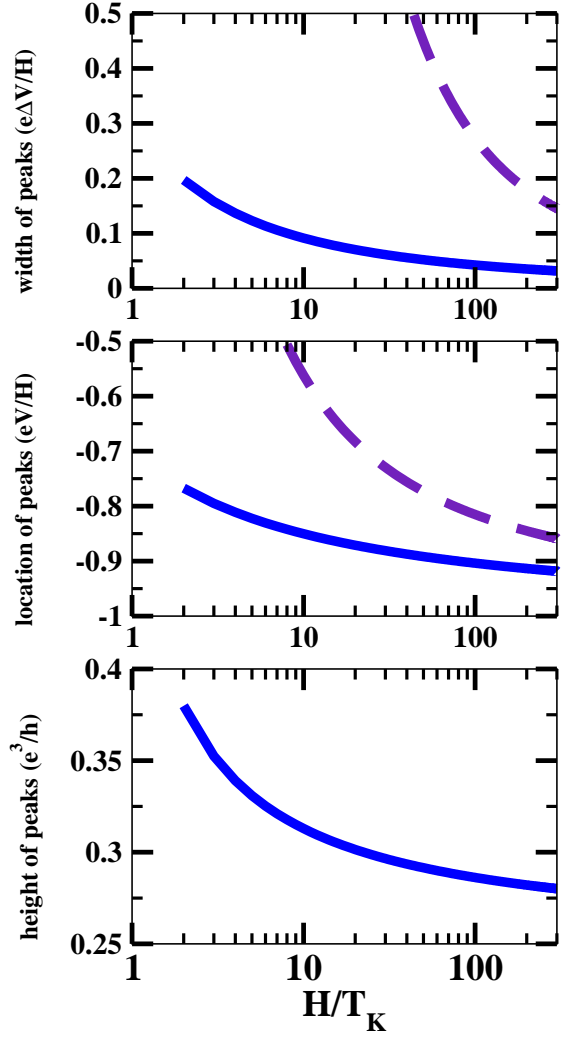}
\end{center}
\caption{Plots describing the evolution
of the differential noise peaks with increasing magnetic field.
In the top panel are plots of the widths of the two peaks, in
the middle panel plots of the two peaks' locations, and 
in the
bottom panel, a single plot of both peaks' (identical) height.}
\end{figure}

\noindent These two phases occur 
at energies parameterized by values of k given by 
\begin{equation}\label{eVxiv}
g(k) = I^{-1} + \frac{1}{2} \tan (\frac{3\pi}{8} / \frac{\pi}{8}).
\end{equation}
Hence the width of this peak is
\begin{eqnarray}\label{eVxv}
e\Delta V &=& \frac{H}{2\pi}\bigg(\tan^{-1} (\frac{1}{I^{-1} - b + 
\frac{1}{2}\tan (\frac{\pi}{8})})\cr\cr
&& ~~~~~~~~- \tan^{-1} (\frac{1}{I^{-1} - b + \frac{1}{2}\tan (\frac{3\pi}{8})})\bigg).
\end{eqnarray}
Similarly, for the peak corresponding to $\delta^{peak\up}_{ho} = \frac{3\pi}{2}$,
the phases of the half-maxima are
$$
\delta^{FWHM\up}_{ho} = \frac{5\pi}{4},~\frac{7\pi}{4},
$$
and consequently, the FWHM of this peak is 
\begin{eqnarray}\label{eVxvi}
e\Delta V &=& \frac{H}{2\pi}\bigg(\tan^{-1} (\frac{1}{I^{-1} - b 
- \frac{1}{2}\tan (\frac{3\pi}{8})})\cr
&& ~~~~~~~~
- \tan^{-1} (\frac{1}{I^{-1} - b - \frac{1}{2}\tan (\frac{\pi}{8})})\bigg).
\end{eqnarray}
The width of the two peaks is thus notably different, with the peak
occurring nearer to $e|V| = H$ the narrower.

The various peak parameters are plotted in Figure 5.  We see in the bottom
panel that the height of the peaks approaches an asymptotic value of
$e^3/4h$ logarithmically in $H/T_k$.   This limit corresponds to a situation
where only spin $\up$ electrons contribute to the current.  In the middle
panel of Figure 5 are plotted the biases, $V_\pm$, at which the two peaks
occur.  In the large $H/T_k$ limit, $eV_\pm$ approaches $H$.  However even
at large $H/T_k$ there is a significant deviation from $H$, a consequence
of the logarithmic dependence upon $H/T_k$.  This behaviour mimics that of
the peak in the differential conductance:  for large $H/T_k$, this peak
occurs at a value of $eV$ distinctly smaller than $H$ \cite{long}.  
Finally in the
top panel of Figure 5
is plotted the width of the two peaks.  Again the peak that
occurs at a bias closer to -$H$ is the more narrow of the two.

\section{Conclusions}
\label{conclusions}

To summarize, we have presented analyzes of the
moments of the current using both Fermi liquid
perturbation theory and the Bethe ansatz.
Both approaches however remain incomplete.  In the Fermi liquid
approach, while we have derived results for the higher moments of
the current valid, we believe, at all orders of the interaction
strength, we have done so only at the lowest non-trivial order in
the voltage.  To expand these computations to higher orders
in the voltage remains a challenging problem for future research.
In terms of the Bethe ansatz treatment, we have managed to identify various
robust features of the current noise that should be experimentally
identifiable in current realizations
of quantum dots.  In particular, we have identified a double peaked structure
in the noise that appears at finite magnetic fields.
But because of our use of {\it equilibrium} scattering amplitudes, we have 
been able to focus only upon qualitative aspects of the physics.  Quantitatively,
the methodology produces results at variance with Fermi liquid theory.
It thus remains an important goal to pinpoint the precise origin of this
disagreement.

\end{document}